\documentclass{article}
\usepackage{spconf,amsmath,graphicx,color}
\usepackage{arydshln}
\usepackage{xcolor}
\usepackage[ruled,vlined,linesnumbered]{algorithm2e}
\usepackage[symbol]{footmisc}

\SetCommentSty{mycommfont}

\title{Investigation of End-to-End Speaker-Attributed ASR for\\Continuous Multi-talker Recordings}
\name{\begin{tabular}{c}Naoyuki Kanda$^1$, Xuankai Chang$^2$\sthanks{Work performed during internship at Microsoft.},
Yashesh Gaur$^1$,  Xiaofei Wang$^1$,
Zhong Meng$^1$,\\ 
Zhuo Chen$^1$, Takuya Yoshioka$^1$\end{tabular}}
\address{$^1$Microsoft Corp., USA\\$^2$Johns Hopkins University, USA}
\begin{document}
\ninept
\maketitle
\begin{abstract}
Recently, an end-to-end (E2E) speaker-attributed automatic speech recognition (SA-ASR) model was proposed as a joint model of speaker counting, speech recognition and speaker identification for monaural overlapped speech. It showed promising results for simulated speech mixtures consisting of various numbers of speakers. However, the model required prior knowledge of speaker profiles to perform speaker identification, which significantly limited the application of the model. In this paper, we extend the prior work by addressing the case where no speaker profile is available. Specifically, we perform speaker counting and clustering by using the internal speaker representations of the E2E SA-ASR model to diarize the utterances of the speakers whose profiles are missing from the speaker inventory. We also propose a simple modification to the reference labels of the E2E SA-ASR training which helps handle continuous multi-talker recordings well. We conduct a comprehensive investigation of the original E2E SA-ASR and the proposed method on the monaural LibriCSS dataset. Compared to the original E2E SA-ASR with relevant speaker profiles, the proposed method achieves a close performance without any prior speaker knowledge. We also show that the source-target attention in the E2E SA-ASR model provides information about the start and end times of the hypotheses.
\end{abstract}

\section{Introduction}
\label{sec:intro}

Speaker-attributed automatic speech recognition (SA-ASR), 
which recognizes "who spoke what", is essential to meeting transcription.
SA-ASR requires to count the number of speakers,
transcribe the utterances, and identify or diarize the speaker of each utterance
from conversational recordings where some utterances are usually overlapped.
It has
a long research history, 
from the projects in the early 2000's
\cite{fiscus2007rich,janin2003icsi,carletta2005ami}
to the recent international efforts such as
the CHiME \cite{barker2018fifth,watanabe2020chime} and DIHARD  \cite{ryant2018first,ryant2019second} challenges. 
While significant progress has been made  especially 
in multi-microphone settings~(e.g., \cite{kanda2018hitachi,yoshioka2019advances,kanda2019guided,medennikov2020stc}), 
SA-ASR for monaural audio remains  challenging due to the difficulty
in handling overlapped speech for both ASR and speaker diarization/identification.

One dominant approach to SA-ASR is applying speech separation (e.g., \cite{hershey2016deep,chen2017deep,yu2017permutation})
before ASR and speaker diarization/iden-tification. 
However, a speech separation module is often designed and trained with
a signal-level criterion and therefore suboptimal for 
the downstream 
modules. 
To overcome this problem, 
joint modeling of multiple modules has been investigated from a variety of
view points.
For example, a number of studies have 
investigated
 joint modeling of 
speech separation and ASR
(e.g., \cite{yu2017recognizing,seki2018purely,chang2019end,chang2019mimo,kanda2019acoustic,kanda2019auxiliary}).
Several methods were also proposed for integrating speaker identification and speech separation~\cite{wang2019speech,von2019all,kinoshita2020tackling}. 
A few studies attempted to improve the speaker diarization
by leveraging ASR results~\cite{park2018multimodal,park2020speaker}.

However, only a limited number of research works investigated the 
joint modeling of all necessary modules of SA-ASR.
\cite{el2019joint} proposed to generate transcriptions for
different speakers interleaved by speaker role tags 
to recognize doctor-patient conversations based on
a recurrent neural network transducer (RNN-T). 
Although promising results were shown,
the method cannot deal with speech overlaps
due to the monotonicity constraint of RNN-T. 
Furthermore,
their method is difficult to extend to  
an arbitrary number of speakers
because the target speaker roles need to be uniquely defined. 
In \cite{mao2020speech}, the authors applied a 
similar technique to \cite{el2019joint}
by interleaving multiple utterances with 
speaker identity tags instead of speaker role tags.
To handle speakers who were unseen in the training data,
the authors used speaker identity tags from the training data even for
the unseen test speakers,
or they simply applied a separated speaker diarization module.
However, their method showed severe degradation of 
ASR and speaker diarization accuracy
when
the oracle utterance boundaries were not used.
\cite{kanda2019simultaneous} proposed a
joint decoding framework for
overlapped speech recognition and speaker diarization,
where 
speaker embedding estimation and target-speaker ASR were performed 
alternately.
While their formulation is applicable to any number of speakers, 
the method was actually implemented and evaluated
in a way that could be used only for the two-speaker case, as target-speaker ASR was performed with an auxiliary output branch
representing a single interference speaker \cite{kanda2019auxiliary}.

Recently, an end-to-end (E2E) SA-ASR model has been proposed
as a joint model of speaker counting, speech recognition, 
and speaker identification for monaural (possibly) overlapped speech \cite{kanda2020joint}.
It was trained to maximize 
the joint probability for multi-talker speech recognition and speaker identification,
and achieved a significantly lower 
speaker-attributed word error rate (SA-WER) than a system
 that separately performs overlapped speech recognition and speaker
identification.
However, 
the model only works with
a speaker inventory
that includes
the profiles (i.e., embeddings) of all speakers involved in the input speech.
This requirement strongly limited its application to real scenarios.

In this paper, 
we  extend  the previous
E2E SA-ASR work
to address the case where no speaker profile is available.
Specifically, we propose to cluster the internal speaker
representations of the E2E SA-ASR model
to diarize the utterances of 
the speakers whose speaker profiles are not included in the speaker inventory.
Combined with a silence-region detector, 
this also allows a very long-form signal spanning an entire meeting to be handled. 
We also propose a simple modification to the reference label construction for the E2E SA-ASR training 
to handle continuous multi-talker recordings more effectively. 
Comprehensive experimental results using the monaural LibriCSS dataset \cite{chen2020continuous}, consisting of eight-speaker sessions, 
show the effectiveness of the proposed method.

\section{Review: E2E SA-ASR}
\label{sec:e2e-sa-asr}
\subsection{Overview}

In this section, we review the E2E SA-ASR method
proposed in \cite{kanda2020joint}.
The goal of this method  
 is 
 to estimate a
multi-speaker transcription $Y=\{y_1,...,y_N\}$
and
the speaker identity of each token $S=\{s_1,...,s_N\}$ 
given
acoustic input 
$X=\{x_1,...,x_T\}$
 and a speaker inventory $\mathcal{D}=\{d_1,...,d_K\}$.
 Here,
$N$ 
is the number of the output tokens,
$T$ is the number of the input frames,
and $K$ is the number of the speaker profiles 
(e.g., d-vector \cite{variani2014deep}) 
in the inventory $\mathcal{D}$.
Following the idea of 
serialized output training (SOT) \cite{kanda2020sot}, 
the multi-speaker transcription $Y$ is represented by concatenating 
individual speakers' transcriptions
interleaved by a special symbol 
$\langle sc\rangle$
representing the speaker change.

In the E2E SA-ASR modeling,
it is assumed that
 the profiles of all
the speakers involved in the input speech 
are included in $\mathcal{D}$. 
Note that, as long as this assumption holds,
the speaker inventory may include 
 irrelevant speakers' profiles.

\begin{figure}[t]
  \centering
  \includegraphics[width=\linewidth]{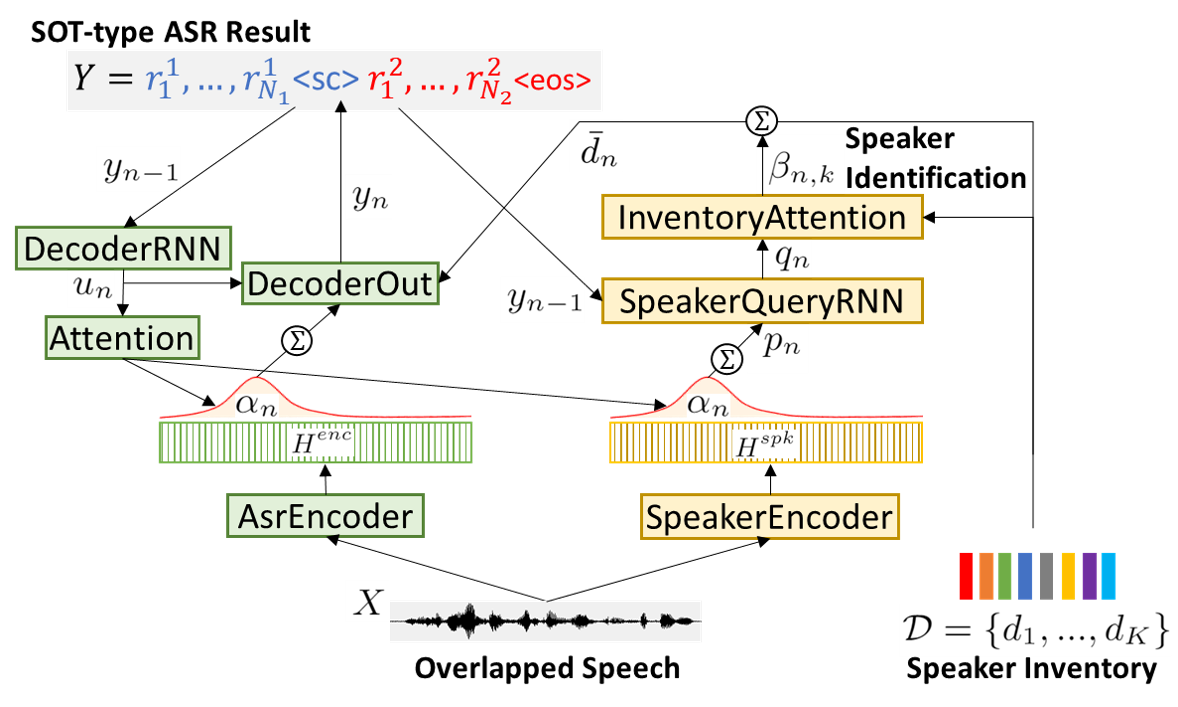}
  \vspace{-7mm}
  \caption{E2E SA-ASR model.}
  \label{fig:asr_models}
  \vspace{-5mm}
\end{figure}

\subsection{Model architecture}

Figure \ref{fig:asr_models} shows 
the architecture of the E2E SA-ASR model.
It consists of  
ASR-related blocks (shown in green), and speaker identification-related
blocks (shown in yellow).
The computation consists of the following five steps. 

\subsubsection{Step1: applying ASR- and speaker- encoders}
Given the acoustic input $X$, 
an ASR encoder firstly converts $X$ 
into a sequence, $H^{enc}$, of embeddings for ASR, i.e.,  
 \begin{align}
 H^{enc} &=\{h^{enc}_1,...,h^{enc}_T\}={\rm AsrEncoder}(X).  \label{eq:enc} 
\end{align}
At the same time,
a speaker encoder converts $X$ into 
a sequence, $H^{spk}$, of 
embeddings representing the speaker features of
the input $X$ as follows:
 \begin{align}
 H^{spk} &= \{h^{spk}_1,...,h^{spk}_T\} = {\rm SpeakerEncoder}(X).  \label{eq:spkenc}
\end{align}

\subsubsection{Step2:  attention weight estimation}
Secondly, at each decoder step $n$, an attention module 
generates attention weight $\alpha_n =\{\alpha_{n,1},...,\alpha_{n,T}\}$
as
\begin{align}
 \alpha_n &= {\rm Attention}(u_n, \alpha_{n-1}, H^{enc}), \label{eq:att} \\
 u_n &={\rm DecoderRNN}(y_{n-1}, c_{n-1}, u_{n-1}),\label{eq:att2} 
\end{align}
where 
$u_n$ is a decoder state vector
at the $n$-th step, and $c_{n-1}$ is a context vector at the previous time step.

\subsubsection{Step3: calculating context vector for ASR}
Then, context vector $c_n$ for the current decoder step $n$ is generated as a weighted sum of
the encoder embeddings as follows:
\begin{align}
 c_n&=\sum_{t=1}^T \alpha_{n,t}h^{enc}_t.
 \label{eq:context}
\end{align}

\subsubsection{Step4: speaker identification}
At every decoder step $n$,
the attention weight $\alpha_n$ is also applied to $H^{spk}$
to extract an attention-weighted average, $p_n$, of the speaker embeddings as 
\begin{align}
p_n
    &=\sum_{t=1}^T \alpha_{n,t}h^{spk}_t.
\end{align}
Note that 
$p_n$
could be contaminated by interfering speech
because some time frames include two or more speakers.

The speaker query RNN in Fig.~\ref{fig:asr_models} then generates a
speaker query $q_n$ given the speaker embedding $p_n$,
the previous output $y_{n-1}$, and the previous 
speaker query $q_{n-1}$, i.e., 
\begin{align}
q_n &= {\rm SpeakerQueryRNN}(p_n,y_{n-1},q_{n-1}).
\end{align}
With the speaker query $q_n$,
an attention module for speaker inventory (shown as InventoryAttention in the diagram) 
estimates
attention weight $\beta_{n,k}$ for each profile in $\mathcal{D}$:
\begin{align}
b_{n,k}&=\frac{q_n\cdot d_k}{|q_n||d_k|}, \label{eq:sp_att_cos}\\
\beta_{n,k}&= \frac{\exp(b_{n,k})}{\sum_j^K \exp(b_{n,j})}.  
\label{eq:sp_att_softmax}
\end{align}
The attention weight $\beta_{n,k}$ can be seen as 
a posterior probability of person $k$ speaking the $n$-th token
given all the previous tokens and speakers as well as $X$ and $\mathcal{D}$, i.e., 
\begin{align}
Pr(s_n=k|y_{1:n-1},s_{1:n-1},X,\mathcal{D}) \sim\beta_{n,k}. \label{eq:spk-prob}
\end{align}

Attention-weighted 
speaker profile $\bar{d}_n$ is also calculated
based on the attention weight $\beta_{n,k}$ and input profile $d_k$ as
\begin{align}
\bar{d}_n=\sum_{k=1}^{K}\beta_{n,k}d_k. \label{eq:weighted_prof}
\end{align}

\subsubsection{Step5: ASR using context  and speaker vectors}
Finally, 
the output distribution for $y_n$ 
is estimated given
the context vector $c_n$,
the decoder state vector $u_n$,
and 
the weighted speaker vector $\bar{d}_n$
as follows: 
\begin{align}
Pr(y_n|y_{1:n-1},s_{1:n},X,\mathcal{D}) &\sim {\rm DecoderOut}(c_n,u_n,\bar{d}_n) \nonumber \\
&\hspace{-30mm}={\rm Softmax}(W_{out}\cdot {\rm LSTM}(c_n+u_n+W_d \bar{d}_n)). \label{eq:token-prob} 
\end{align}
Here,
it is assumed
that $c_n$ and $u_n$ have the same dimensionality,
and $W_d$ is a matrix to change the dimension of $\bar{d}_n$ to
that of $c_n$.
Variable $W_{out}$ is the affine transformation matrix of the final layer.
Typically, 
${\rm DecoderOut}$ consists of
a single affine transform with a softmax output layer.
However, in this work, we insert one LSTM just before the affine transform
as it improves 
the efficacy of the SOT model as shown in \cite{kanda2020sot}.

\subsection{Training}
All  network parameters are
optimized by maximizing
the speaker-attributed maximum mutual information criterion as follows: 
\begin{align}
\mathcal{F}^{\mathrm{SA-MMI}}&=\log Pr(Y,S|X,\mathcal{D}) \label{eq:samll-1}\\
&\hspace{-10mm}= \log \prod_{n=1}^{N}\{Pr(y_{n}|y_{1:n-1}, s_{1:n}, X, \mathcal{D}) \nonumber \\ 
&\hspace{-10mm} \;\;\;\;\;\;\;\;\;\;\;\;\;\;\; \cdot Pr(s_{n}|y_{1:n-1}, s_{1:n-1}, X, \mathcal{D})^\gamma \}. \label{eq:samll-2}
\end{align}
Here,  
$\gamma$ is a scaling parameter 
for the speaker estimation probability and is set to 0.1 per \cite{kanda2020joint}.

\subsection{Decoding}
An extended beam search algorithm is used for decoding for the E2E SA-ASR.
With the conventional beam search,
each hypothesis 
contains estimated tokens accompanied by the posterior probability of 
the hypothesis.
In addition to these,
a hypothesis for the E2E SA-ASR method contains
speaker estimation $\beta_{n,k}$.
 Each hypothesis expands until $\langle eos\rangle$ is detected,
 and the estimated tokens in each hypothesis are grouped by  
 $\langle sc\rangle$ 
 to form multiple utterances.
 For each utterance,
the speaker with the highest $\beta_{n,k}$
value at the point of  $\langle sc\rangle$ or  $\langle eos\rangle$ token is
 selected as the predicted speaker of that utterance \footnote{
We observed slight performance improvement 
by using the
 speaker estimation at the end of an utterance (i.e., the $\langle sc\rangle$ or  $\langle eos\rangle$ position) instead of  the 
 original scheme proposed in \cite{kanda2020joint} which uses the average $\beta_{n,k}$ values calculated over all tokens  
of the utterance.}
 Finally, when the same speaker is predicted for multiple
 utterances, those utterances are concatenated to form a single utterance.

\section{Extensions of E2E SA-ASR}
This section describes 
our proposed extensions of the E2E SA-ASR
for recognizing continuous multi-talker recordings
without prior speaker knowledge.

\subsection{Combination of E2E SA-ASR and speaker clustering}

The E2E SA-ASR 
 requires the speaker inventory to include 
the  profiles 
of all speakers involved in the input speech.
However, 
it is often difficult to prepare
such a speaker inventory 
 for various reasons, including 
 the participation of  
 guest speakers who are not originally invited to a meeting and the privacy concern about voice enrollment.

To cope with the case where no prior speaker knowledge is available, 
we combine the E2E SA-ASR and 
speaker clustering.
Here, we assume we have
 a well-trained E2E SA-ASR model.
Then, our proposed procedure 
to recognize long audio recordings
is as follows.
\begin{enumerate}
    \item Firstly, we apply a silence-region detector to divide an input long audio 
    recording
    into multiple shorter segments at every silence regions. Each segment may include multiple utterances of different speakers with overlaps.
    \item Then, we apply the E2E SA-ASR for each segment with a set of example speaker profiles who do not appear in the input audio.
    \item Finally, we cluster the speaker query vectors $q_n$ of the recognized hypotheses (i.e., the query vectors obtained at the last token of each utterance) to count and diarize the speakers. 
    Specifically, we first determine the number of clusters 
    based on normalized maximum eigengap (NME) \cite{park2019auto}, 
    and then perform spectral clustering 
    with a normalized graph 
    Laplacian matrix \cite{von2007tutorial}. \footnote{
In \cite{park2019auto}, 
spectral clustering was applied
to a binarized and unnormalized graph Laplacian matrix
after speaker counting.
However, we applied the conventional spectral clustering
with a normalized graph Laplacian as this yielded  slightly better results in our preliminary experiments.}
\end{enumerate}

One may have multiple questions about this procedure.
For example, how many example (irrelevant) speaker profiles
are necessary in step 2? How does the silence-region detector in
step 1 affect  the
final result?
How about 
using 
the weighted profile $\bar{d}_n$ for 
speaker counting and clustering in step 3 instead of the speaker query $q_n$?
We will 
experimentally examine these questions
in Section \ref{sec:experiments}.

\begin{figure}[t]
  \centering
  \includegraphics[width=\linewidth]{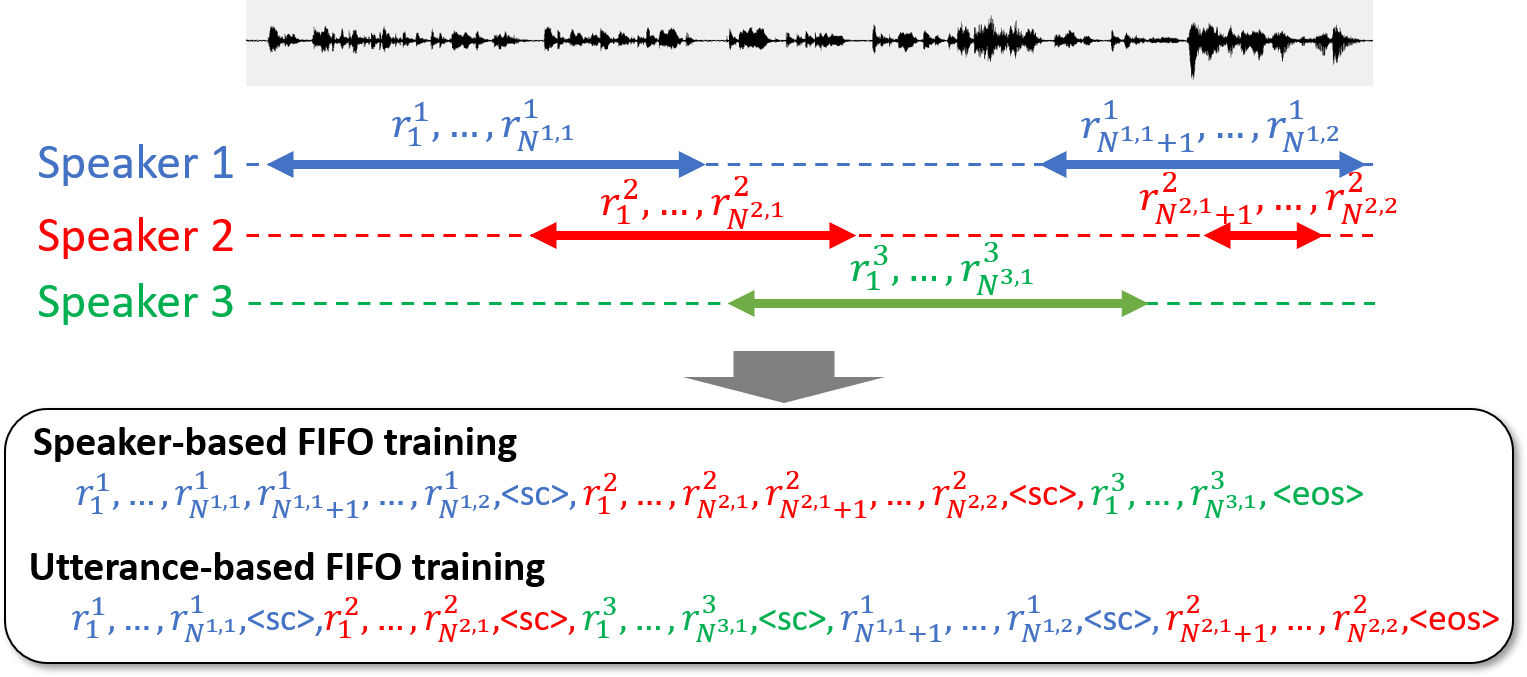}
  \vspace{-7mm}
  \caption{Speaker-based and utterance-based FIFO training.}
  \label{fig:sot_labels}
  \vspace{-5mm}
\end{figure}

\subsection{Modified FIFO training}
\label{sec:fifo_order}

We also introduce a simple yet effective modification of
the reference transcription construction for the E2E SA-ASR training.
In the previous work \cite{kanda2020joint},
the authors trained the E2E SA-ASR model with 
overlapped speech of 
up to three utterances.
However, in real conversation,
there are many cases where the same speaker
utters multiple times in one continuous audio segment 
as illustrated in the upper part of Fig. \ref{fig:sot_labels}.
In this example, three people are speaking in one audio segment,
and $r^i_j$ represents the $j$-th reference token of speaker $i$.
The term $N^{i,u}$ represents the end position of $u$-th utterance of
speaker $i$.

The previous work \cite{kanda2020joint}
employed the 
first-in first-out (FIFO) training scheme
\cite{kanda2020sot}, where
the reference labels of different speakers are sorted
by their start times and concatenated
by $\langle sc\rangle$ token. 
Since the $\langle sc\rangle$ token represents the 
{\it speaker} change, 
the transcriptions of individual speakers  
are sorted by the times they start speaking. 
We call this original version {\it speaker-based} FIFO training,
and shows an example in Fig. \ref{fig:sot_labels}.

Alternatively, we may sort the reference labels
according to the start time of {\it each utterance}
and join the utterances with the $\langle sc\rangle$ token. 
Note that this scheme implicitly assumes that we can define what an end of an utterance is in continuous speech. 
We call this modified version {\it utterance-based} FIFO training,
as illustrated in Fig. \ref{fig:sot_labels}.
In the next section, we experimentally investigate which FIFO training scheme results in better performance.

\begin{table}[t]
  \caption{CpWERs (\%) of the E2E SA-ASR 
  with speaker inventory of 8 relevant speakers.
  LSTM-LM was not used in this experiment.
  Audio recordings were segmented at non-speech points based on  oracle
  boundary information. }
  \label{tab:sot-label}
  \centering
  {\footnotesize
  \begin{tabular}{c|cccccc|c}
    \hline
FIFO & \multicolumn{7}{c}{cpWER (\%) for different overlap ratio} \\ 
order&  0S & 0L & 10 & 20 & 30 & 40 & Avg. \\ \hline
Speaker  & 7.1  & 6.8 & 21.4 & 24.3 & 42.7 & 44.6 & 26.7 \\ 
Utterance & 6.9 & 7.0 & 11.2 & 15.0 & 28.4 & 30.3 & 17.8 \\ \hline
  \end{tabular}
  }
  \vspace{-3mm}
\end{table}

\setlength{\dashlinedash}{2pt}
\setlength{\dashlinegap}{2pt}
\begin{table*}[t]
  \caption{CpWERs (\%) and speaker counting errors with different speaker profile settings. 
  The audio recordings were segmented at non-speech points based on oracle
  boundary information. 
  LSTM-LM was used in this experiment.}
  \label{tab:summary}
  \centering
  {\footnotesize
  \begin{tabular}{cccc|cccccc|c|c}
    \hline
\# of relevant   & \# of irrelevant      & Speaker   & Speaker &    \multicolumn{7}{c|}{cpWER (\%) for different overlap ratio} & Mean speaker \\ 
profiles    & profiles & clustering  & counting &  0S & 0L & 10 & 20 & 30 & 40 & {\bf Avg.} & counting error\\ \hline 
\multicolumn{4}{l|}{\it E2E SA-ASR} & \multicolumn{6}{c|}{}& \\
 8 & 0 &-  & automatic  &  6.1 & 5.7 & 9.3 & 15.3 & 26.7 & 30.3 & {\bf 16.9} & 0.00 \\ 
  8 & 5 & - & automatic & 6.2 & 5.8 & 10.1 & 14.9 & 26.0 & 36.1 & {\bf 18.0} & 0.37 \\ 
 8 & 10 & - & automatic & 6.2  & 6.2 & 9.9 & 15.4 & 27.1 & 36.7 & {\bf 18.5} & 0.91 \\ 
 8 & 20 &-  & automatic & 7.0  & 6.9 & 10.4 & 15.8 & 27.6 & 37.4 & {\bf 19.0} & 1.59 \\ 
 8 & 100 & - & automatic  & 10.2  & 10.2 & 12.1 & 17.3 & 31.5 & 41.9 & {\bf 22.1} & 3.91 \\ \hdashline
 0 & 10 & - & automatic & 71.3  & 67.7 & 64.4 & 67.8 & 80.0 & 78.1 & {\bf 72.1} & 1.19 \\ 
 0 & 20 & - & automatic  & 64.1  & 61.2 & 68.8  & 71.3 & 84.1  & 78.5 & {\bf 72.5} & 7.17  \\ 
 0 & 100 & - & automatic & 69.4 & 73.3 & 78.3 & 83.0 & 90.2 & 87.5 & {\bf 81.3} & 20.09 \\ \hline 
\multicolumn{4}{l|}{\it E2E SA-ASR + Speaker Clustering (proposed method)} & \multicolumn{6}{c|}{} & \\
0 & 100    & $\surd$  &  oracle  & 5.6 & 6.8 & 9.3 & 14.2 & 26.4 & 30.3 & {\bf 16.7}& 0.00 \\ 
0 & 100    & $\surd$ & NME (max=8) & 6.6 & 9.0  &  13.3 &  14.2 & 26.4 & 30.7 & {\bf 17.9} & 0.11 \\  
0 & 100    & $\surd$ & NME (max=12) & 6.6 & 13.7  & 14.9 & 15.9 & 28.1 & 30.7 & {\bf 19.3} & 0.30 \\ 
0 & 100    & $\surd$ & NME (max=16) & 11.0 & 13.7  & 14.9 & 15.9 & 28.1 & 30.7 & {\bf 20.0} & 0.43 \\  \hline
  \end{tabular}
  }
  \vspace{-5mm}
\end{table*}

\section{Experiments}
\label{sec:experiments}
\subsection{Evaluation settings}
\subsubsection{Evaluation data}
We evaluated the effectiveness of the proposed method by 
using the LibriCSS dataset \cite{chen2020continuous},
which comprises conversation-like recordings 
created based on the LibriSpeech corpus \cite{panayotov2015librispeech}.
The dataset consists of 10 hours of recordings of concatenated LibriSpeech utterances that were played back by multiple loudspeakers 
in a meeting room and 
captured by a seven-channel microphone array. 
While the recordings have seven channels, 
we used only the first channel data (i.e. monaural audio)
for all our experiments.

The LibriCSS dataset consists of 
10 sessions, each 
being one hour long and comprising eight speakers. 
Per \cite{chen2020continuous}, 
each session is decomposed to six 10-minute-long 
``mini-sessions'' that have different overlap ratios 
ranging from 0\% to 40\%. 
The recordings of the first session (Session 0) was used to tune the decoding parameters,
and those in the rest of 9 sessions (Session 1--9) were used for the evaluation.
Note that there are two types of mini-sessions for the 0\% overlap case: 
one has only 0.1-0.5 sec 
of silence between adjacent utterances (called ``0S''); one has 2.9-3.0 sec of silence between the adjacent utterances (called ``0L'').

\subsubsection{Training data}

For the E2E SA-ASR training, 
we used
multi-speaker signals that were generated by room simulation from 
the 960 hours of LibriSpeech training data (``train\_960'')
\cite{panayotov2015librispeech,povey2011kaldi}.
We generated 500,000 training samples,
each of which was a 
mixture of multiple utterances randomly selected
from train\_960.
When the utterances were mixed,
each utterance was shifted by a random delay
to simulate partially overlapped conversational recordings.
Each training sample was generated 
under the following conditions.
\begin{itemize}
\setlength{\itemsep}{0pt}
\setlength{\parskip}{0pt}
    \item The number of speakers was randomly chosen from 1 to 5.
    \item The number of utterances was randomly chosen from 1 to 5.
    \item The start times of different utterances were apart by 0.5 sec or longer.
    \item Every utterance in each mixed audio sample had at least 
    one speaker-overlapped region with other utterances.
    \item Utterances of the same speakers do not overlap.
\end{itemize}
Before mixing the source utterances, 
a room impulse response generated by 
 the image method was applied to each utterance \cite{ko2017study}.
In addition,
random noise was generated by following
 \cite{habets2007generating}, and added at a random SNR from 10 to 40 dB
 after mixing the utterances. 
Finally, the volume of the mixed audio was changed by a random scale between 0.125 and 2.0.

In addition to the multi-speaker signals,
speaker profiles were generated for each training sample as follows. 
For a training sample consisting of $S$ speakers, the number of the profiles was randomly selected 
from $S$
to  8. 
Among those profiles, $S$ profiles were 
for the speakers involved in the overlapped speech. 
The utterances for creating the profiles of these speakers were 
different from those constituting the input overlapped speech. 
The rest of the profiles were randomly extracted from different speakers in train\_960. 
Each profile was extracted by using 
10 utterances.

\subsubsection{Evaluation metric}

The main evaluation metric used 
in this paper is
 the concatenated minimum-permutation
word error rate (cpWER) \cite{watanabe2020chime}.
The cpWER is computed as follows:
(i) concatenate all reference transcriptions for each speaker; (ii) concatenate all hypothesis transcriptions for each detected speaker; 
(iii) compute the WER between the reference and hypothesis and repeat this for all possible
speaker permutations; 
and (iv) pick the lowest WER among them.
The cpWER is affected by both
the speech recognition and speaker diarization results.

Besides cpWER, 
we evaluated the mean
speaker counting error,
which is the
absolute difference between
the estimated number of speakers 
and the actual number of speakers 
(= 8 in 
LibriCSS)
averaged 
over all mini-sessions.
We also analyzed the source-target attention of our system in terms of the diarization error rate (DER). 
It should be noted that the 
mean speaker counting error and the DER 
are not the performance metrics we care, and they were evaluated only for analysis purposes. 
The hyper-parameters of our systems were tuned on the development set to improve only the cpWER.

\subsubsection{Model settings}
In our experiments,
an 80-dim log mel filterbank extracted every 10 msec was used
for the input feature.
3 frames of features were stacked, and the model was applied 
on top of the stacked features.
For the speaker profile, we used
a 128-dim d-vector \cite{variani2014deep}, 
whose extractor was separately 
trained on VoxCeleb Corpus \cite{nagrani2017voxceleb,chung2018voxceleb2}.
The d-vector extractor 
consisted of 17 convolution layers followed by an average pooling layer, which was a modified version of the one presented in \cite{zhou2019cnn}. 

The AsrEncoder consisted of
 5 layers of 1024-dim 
bidirectional long short-term memory (BLSTM), interleaved with layer normalization~\cite{ba2016layer}.
The DecoderRNN consisted of 
2 layers of 1024-dim unidirectional LSTM,
and the DecoderOut consisted of 1 layer of 1024-dim unidirectional LSTM. 
We used a conventional location-aware content-based attention \cite{chorowski2015attention} with 
a single attention head. 
The SpeakerEncoder had the same architecture as the d-vector extractor except for not having the final average pooling layer.
Our SpeakerQueryRNN consisted of 1 layer of 512-dim unidirectional LSTM.
We used 16k subwords based on a unigram language model \cite{kudo2018subword}
as a recognition unit.

When we trained the E2E SA-ASR model,
we initialized the parameters of
AsrEncoder, Attention, DecoderRNN, and DecoderOut
by the parameter values of a three-speaker SOT-ASR model
trained on  
simulated LibriSpeech utterance mixtures.
We followed the setting 
described in \cite{kanda2020sot} for  pre-training  
the SOT model,
and used the parameter values obtained after 640k training iterations.
We also initialized the SpeakerEncoder parameters 
by using those of the d-vector extractor. 
After the initialization,
we updated the entire network based on $\mathcal{F}^{\mathrm{SA-MMI}}$ 
with $\gamma=0.1$
by using
an Adam optimizer
with a learning rate of 0.00002.
We used 8 GPUs, each of which worked
on 6k frames of minibatch.
We report the results 
of the dev\_clean-based best models found after 120k training iterations. 

In addition to the E2E SA-ASR model described above,
we trained an external language model (LM) that consisted of 4 layers of 2,048-dim LSTM.
As training data,
we generated a text corpus
by (1) shuffling 
the official training text corpus for LibriSpeech and the transcription of train\_960,
and (2) concatenating every consecutive $rand(1,5)$ utterances interleaved by $\langle sc\rangle$ token.
We used the shallow fusion (i.e. simple weighted sum) to combine the E2E SA-ASR and the LM scores
with an LM weight calibrated by using the development set.

\begin{table}[t]
  \caption{Average cpWER (\%) with different numbers of irrelevant profiles (i.e., example profiles) for proposed method using oracle speaker numbers. 
Oracle boundary-based segmentation was used.}
  \label{tab:irrelevant}
  \centering
  {\footnotesize
  \begin{tabular}{c|cccc}
    \hline
 LM & \multicolumn{4}{c}{\# of irrelevant profiles} \\ 
 &           1 & 5 & 10 & 100 \\ \hline
& 19.4 & 19.2 & {\bf 18.9} & {\bf 18.9} \\
$\surd$ & 16.9 & 16.9 & 16.8 & {\bf 16.7} \\ \hline
\end{tabular}
  }
  \vspace{-5mm}
\end{table}

\begin{table}[t]
  \caption{
 CpWERs (\%) with
 different internal speaker
 embeddings for speaker clustering.
Oracle boundary-based segmentation was used.}
  \label{tab:embeddings}
  \centering
  {\scriptsize
  \begin{tabular}{c|cccccc|c}
    \hline
Speaker embedding & \multicolumn{7}{c}{Overlap ratio in \%} \\ 
for clustering & 0S & 0L & 10 & 20 & 30 & 40 & Avg. \\ \hline
Weighted profile {\tiny $\bar{d}_n$}& 10.1 & 10.4 & 17.2 & 23.6 & 34.3  & 34.7 & 23.2 \\ 
Speaker query $q_n$  &  5.6 & 6.8  & 9.3  &  14.2 & 26.4  & 30.3 & 16.7\\ \hline
  \end{tabular}
  }
  \vspace{-3mm}
\end{table}

\begin{table*}[t]
  \caption{CpWERs (\%) with automatic silence-region detector to segment the 
  audio recordings.}
  \label{tab:audo_summary}
  \centering
{\footnotesize
  \begin{tabular}{c|cccc|cccccc|c|c}
    \hline
System &\# of relevant   & \# of irrelevant   & Speaker      & Speaker &    \multicolumn{7}{c|}{cpWER (\%) for different overlap ratio} & Mean speaker\\ 
number & profiles    & profiles & clustering  & counting &  0S & 0L & 10 & 20 & 30 & 40 & {\bf Avg.} & counting error \\ \hline
& \multicolumn{4}{l|}{\it E2E SA-ASR} & \multicolumn{6}{c|}{} & \\
1 & 8 & 0  &- & automatic & 15.7  & 8.0 & 12.5 & 17.5 & 24.3 & 27.6 & {\bf 18.6} & 0.00 \\ 
2 & 8 & 5  &- & automatic & 16.4 & 8.9& 13.4 & 18.2 & 25.0 & 28.3 & {\bf 19.3} & 0.89 \\ 
3 &  8 & 10  &- & automatic & 16.6 & 9.0 & 13.9 & 19.0 & 25.8 & 28.8 & {\bf 19.9} & 1.78  \\ 
4 & 8 & 20  &- & automatic & 18.1 & 10.2 & 14.9 & 19.5 & 27.0 & 30.6 & {\bf 21.1} & 3.11 \\ 
5 &  8 & 100  &- & automatic & 24.0 & 15.5 & 18.8 & 23.8 & 32.3 & 35.4 & {\bf 26.0} & 9.00 \\ \hline
& \multicolumn{4}{l|}{\it E2E SA-ASR + Speaker Clustering (proposed method)} & \multicolumn{6}{c|}{} & \\
6 & 0 & 100  & $\surd$  &  oracle  & 15.8 & 10.3 & 13.4  & 17.1 & 24.4 & 28.6 & {\bf 19.2} & 0.00 \\ 
7 & 0 & 100    &  $\surd$  & NME (max=16) & 24.4 & 12.2 & 15.0  & 17.1 & 28.6  & 28.6 & {\bf 21.8} & 0.31 \\  \hline 
  \end{tabular}
  }
  \vspace{-5mm}
\end{table*}

\begin{table}[t]
  \caption{
Analysis on the source-target attention of two systems in Table \ref{tab:audo_summary} based on DERs (\%).}
  \label{tab:der}
  \centering
  {\scriptsize
  \begin{tabular}{c|cccccc|c}
    \hline
 & \multicolumn{7}{c}{DER (\%) for different overlap ratio } \\ 
 & 0S & 0L & 10 & 20 & 30 & 40 & {\bf Avg.} \\ \hline
System 1 & 15.72 & 11.15 & 12.64 & 14.50 & 18.04 & 17.33 & \hspace{1mm}{\bf 15.23}$^\dagger$ \\ 
System 7 & 19.71 & 12.97 & 13.81 & 14.49 & 19.98  & 17.97 & \hspace{1mm}{\bf 16.75}$^\ddagger$ \\ \hline
  \end{tabular}\\
  $^\dagger$ Miss = 4.72\%, false alarm = 7.04\%, speaker error = 3.47\%\vspace{-1mm}\\
  \hspace{1.5mm}$^\ddagger$ Miss = 4.75\%, false alarm = 7.00\%, speaker error = 5.00\%
  }
  \vspace{-5mm}
\end{table}

\subsection{Evaluation with oracle silence boundary}

We firstly evaluated the proposed method 
with an oracle silence-region detector.
Namely, we divided each recording
at every silence position
obtained from the oracle utterance boundary information.
Note that each segmented audio still consisted of
multiple overlapped utterances of different speakers. 
The minimum and maximum numbers of utterances were found to be 1 and 24, respectively. 
In this subsection,
we used the oracle silence detection. 
The performance using an automatic silence detector is reported in the next subsection.

\subsubsection{Baseline results of E2E SA-ASR}

As a baseline, we evaluated the E2E SA-ASR with
a speaker inventory consisting only of the eight relevant speakers.  
Each speaker's profile was 
extracted by using 5 utterances that were not
included in the recording used for the evaluation.
We firstly compared the speaker-based and 
utterance-based FIFO training schemes that we described in Section \ref{sec:fifo_order}.
The result is shown in Table \ref{tab:sot-label}.
We can see that the utterance-based
FIFO training significantly outperformed the 
speaker-based FIFO training.
Therefore, 
we always used the E2E SA-ASR model based on
 the utterance-based FIFO training in the remaining experiments.

Next, we evaluated the accuracy of the E2E SA-ASR
when the speaker inventory included irrelevant speaker profiles.
In this experiment,
irrelevant speakers were randomly chosen
from train\_960 of LibriSpeech,
and a randomly selected one utterance 
was used to extract the speaker profile of each irrelevant speaker.
The result 
 is shown in the first five rows of Table \ref{tab:summary}.
When no irrelevant profiles were included in the speaker inventory,
the E2E SA-ASR achieved the best cpWER of 16.9\%.
The cpWER gradually deteriorated as the addition of
irrelevant profiles, but the system still achieved
22.1\% of cpWER even with 100 irrelevant profiles.

Finally,
to analyze the impact of the speaker profiles,
we also evaluated the E2E SA-ASR with  
no relevant speaker profiles. 
The results of this experiment are shown from the 6th to 8th rows
of Table \ref{tab:summary},
where we provided 10, 20, or 100 irrelevant profiles as an input
while not using any profiles for the relevant speakers.
Speaker diarization was conducted purely based on
the 
speaker identification result for each utterance.
As expected,
    we observed a very high cpWER of 72.1--81.3\%. 
Note that, 
the mean speaker counting error for the 10 irrelevant profile case
was relatively small (= 1.19) 
just because the given (10) and correct (8) numbers
of speakers were close.

\subsubsection{Results of the proposed method}

We then evaluated the proposed procedure of
the combination of the E2E SA-ASR and speaker clustering.
The results are shown in the last two rows of Table \ref{tab:summary}.
In this experiment, we used 100 irrelevant speaker profiles 
as a set of example profiles.
When we 
applied the speaker clustering
with 
the oracle number of speakers,
the proposed method achieved 16.7\% of cpWER, which was even 
better than the best number obtained by the E2E SA-ASR with the relevant speaker
inventory.
This is because spectral clustering can access to the speaker embeddings
of all utterances while the speaker identification inside the E2E SA-ASR
was done by accessing only the information of the single segment. 
When we estimated the number of speakers by using NME with a maximum possible number of speakers of \{8, 12, 16\}, 
the cpWER was slightly degraded to 17.9--20.0\%. Nonetheless, 
it was still as good as the E2E SA-ASR with 10-20 irrelevant profiles.

We also evaluated 
the effect of the number of irrelevant profiles (= example profiles)
for the combination of the E2E SA-ASR and speaker clustering.
The result of this study is shown in Table \ref{tab:irrelevant}.
It can be seen that using 
too few irrelevant profiles resulted in the degradation of cpWER.
It is because we cannot calculate an appropriate weighted profile $\bar{d}_n$
when we have too few profiles, which ends
up with degrading the overall accuracy.
Note that the computational cost of the inventory attention
(Eq. \eqref{eq:sp_att_cos}--\eqref{eq:weighted_prof}) was negligible even with 100 profiles. 
Thus, we used 100 irrelevant speaker profiles in the following experiments
unless otherwise stated.

We also compared clustering using the weighted profile $\bar{d}_n$ and that using 
speaker query $q_n$. 
The results are shown  in Table \ref{tab:embeddings}.
In this experiment, we applied the E2E SA-ASR with
100 irrelevant speaker profiles,
and then applied speaker clustering given the 
oracle number of speakers.
As seen in the table,
the use of the speaker query $q_n$ resulted in significantly better  speaker clustering performance.

\subsection{Evaluation with automatic silence-region detector}

\subsubsection{Result with respect to cpWER}
We finally evaluated the proposed method 
with an automatic silence-region detector.
In this experiment, we applied
the WebRTC Voice Activity Detector\footnote{https://github.com/wiseman/py-webrtcvad}
for each recording, and 
segmented the audio whenever silence regions were detected.

The result with the automatic silence-region detector
is shown in Table \ref{tab:audo_summary}.
The original E2E SA-ASR with the relevant speaker inventory achieved
18.6\% to 26.0\% of cpWER depending on the number of the additional irrelevant profiles.
On the other hand, 
the proposed combination of the E2E SA-ASR and speaker clustering
achieved 19.2\% of cpWER with oracle speaker counting,
and 21.8\% of cpWER with NME-based speaker counting, respectively.

Compared with the case using the oracle silence-region information,
the cpWER was degraded
by 3.1\%. 
Especially, we noticed that ``0S'' setting showed a severe 
cpWER degradation even though the overlap ratio was 0\%.
With ``0S'', there was very short silence (0.1-0.5 sec) 
between adjacent utterances of different speakers.
As a result, 
segments in ``0S''  
often consisted of consecutive 
 speech of multiple speakers.
We observed the E2E SA-ASR 
sometimes misrecognized the speaker change point for such 
speech,
which resulted in the degradation of cpWER.
Note that the speaker change detection
for non-overlapped speech could be
more difficult than that for overlapped speech
because
speech overlaps could be used as 
a clue of speaker change
besides the difference of voice characteristics.

\subsubsection{Analysis of the source-target attention with DER}
We analyzed the source-target attention $\alpha_n$ 
of the E2E SA-ASR. 
We estimated the start and end times
of each utterance
based on $\alpha_n$ as follows 
and calculated the DER accordingly.
\begin{enumerate}
\setlength{\itemsep}{0pt}
\setlength{\parskip}{0pt}
\item For each utterance hypothesis, the attention ($\alpha_n$)-weighted average of the frame indices was calculated for each token other than 
$\langle sc\rangle$ or $\langle eos\rangle$.
\item The minimum frame index $f_{min}$ and the 
maximum frame index $f_{max}$ were calculated. 
\item The start time $T_s$
was defined as $T_s=\mathrm{max}(0, f_{min} \cdot T_{f} - T_{m})$. The 
end time $T_e$ was defined as $T_e=f_{max} \cdot T_{f} + T_{m}$. 
\end{enumerate}
Here, $T_{f}$ is the frame shift in second, and it was 0.03 sec according to our model settings. 
The term $T_{m}$ is a heuristic margin tuned by the development set, 
and it was determined as 0.5 sec in our experiment.

The DER result is shown in Table \ref{tab:der}. 
In this evaluation,
we calculated the DER without a collar margin,
and the overlapping regions were included in the DER calculation.
As shown in the table, 
the E2E SA-ASR systems
showed 15.23--16.75\% of DER on average.
In the high overlap test sets (with the overlap ratios of 20\%--40\%),
the DERs were significantly
better than the overlap ratios of the input audio,
which indicates that the source-target
attention scanned the encoder embeddings 
back and forth 
to recognize 
overlapped 
utterances one by one
as originally designed
by SOT \cite{kanda2020sot}.
On the other hand,
the DER was as high as 11.15\% even for 
the non overlapped speech (0L). 
This could be because 
the our model is 
optimized to achieve good SA-ASR accuracy,
unlike other diarization methods, 
such as the end-to-end neural diarization
\cite{fujita2019end,fujita2019end2} or target-spekaer voice activity
detection \cite{medennikov2020stc,medennikov2020target}, 
that are optimized for DER. 
 That being said, the result shows that the source-target attention in the E2E SA-ASR model
provides information about the start and end times of the hypotheses and thus can be used for applications requiring both the time boundary and the recognition result.

\section{Conclusion}
\label{sec:conclusion}

In this paper, we proposed
to apply speaker counting and clustering to 
the speaker query of an E2E SA-ASR model to diarize utterances 
of speakers whose speaker profiles are not included 
in the speaker inventory.  
We also proposed a simple yet effective modification to the reference label construction
for E2E SA-ASR training, which helps cope with the continuous 
multi-talker recordings.  
In the evaluation, 
compared with the original E2E SA-ASR with a speaker inventory consisting only of relevant speaker profiles, 
 the proposed method achieved a close cpWER
 even without any prior speaker knowledge.

\bibliographystyle{IEEEbib}
\bibliography{mybib}

\begin{thebibliography}{10}

\bibitem{fiscus2007rich}
Jonathan~G Fiscus, Jerome Ajot, and John~S Garofolo,
\newblock ``The rich transcription 2007 meeting recognition evaluation,''
\newblock in {\em Multimodal Technologies for Perception of Humans}, pp.
  373--389. Springer, 2007.

\bibitem{janin2003icsi}
Adam Janin, Don Baron, Jane Edwards, Dan Ellis, David Gelbart, Nelson Morgan,
  Barbara Peskin, Thilo Pfau, Elizabeth Shriberg, Andreas Stolcke, et~al.,
\newblock ``The {ICSI} meeting corpus,''
\newblock in {\em Proc. ICASSP}, 2003, vol.~1, pp. I--I.

\bibitem{carletta2005ami}
Jean Carletta, Simone Ashby, Sebastien Bourban, Mike Flynn, Mael Guillemot,
  Thomas Hain, Jaroslav Kadlec, Vasilis Karaiskos, Wessel Kraaij, Melissa
  Kronenthal, et~al.,
\newblock ``The {AMI} meeting corpus: A pre-announcement,''
\newblock in {\em International workshop on machine learning for multimodal
  interaction}. Springer, 2005, pp. 28--39.

\bibitem{barker2018fifth}
Jon Barker, Shinji Watanabe, Emmanuel Vincent, and Jan Trmal,
\newblock ``The fifth {'CHiME'} speech separation and recognition challenge:
  Dataset, task and baselines,''
\newblock {\em Proc. Interspeech}, pp. 1561--1565, 2018.

\bibitem{watanabe2020chime}
Shinji Watanabe, Michael Mandel, Jon Barker, Emmanuel Vincent, Ashish Arora,
  Xuankai Chang, Sanjeev Khudanpur, Vimal Manohar, Daniel Povey, Desh Raj,
  et~al.,
\newblock ``{CHiME-6} challenge: Tackling multispeaker speech recognition for
  unsegmented recordings,''
\newblock in {\em Proc. CHiME 2020}, 2020.

\bibitem{ryant2018first}
Neville Ryant, Kenneth Churchb, Christopher Cieria, Alejandrina Cristiac, Jun
  Dud, Sriram Ganapathye, and Mark Libermana,
\newblock ``First {DIHARD} challenge evaluation plan,''
\newblock 2018.

\bibitem{ryant2019second}
Neville Ryant, Kenneth Church, Christopher Cieri, Alejandrina Cristia, Jun Du,
  Sriram Ganapathy, and Mark Liberman,
\newblock ``The second {DIHARD} diarization challenge: Dataset, task, and
  baselines,''
\newblock {\em Proc. Interspeech}, pp. 978--982, 2019.

\bibitem{kanda2018hitachi}
Naoyuki Kanda, Rintaro Ikeshita, Shota Horiguchi, Yusuke Fujita, Kenji
  Nagamatsu, Xiaofei Wang, Vimal Manohar, Nelson Enrique~Yalta Soplin, Matthew
  Maciejewski, Szu-Jui Chen, et~al.,
\newblock ``The {Hitachi/JHU CHiME-5} system: Advances in speech recognition
  for everyday home environments using multiple microphone arrays,''
\newblock in {\em Proc. CHiME-5}, 2018, pp. 6--10.

\bibitem{yoshioka2019advances}
Takuya Yoshioka, Igor Abramovski, Cem Aksoylar, Zhuo Chen, Moshe David,
  Dimitrios Dimitriadis, Yifan Gong, Ilya Gurvich, Xuedong Huang, Yan Huang,
  et~al.,
\newblock ``Advances in online audio-visual meeting transcription,''
\newblock in {\em Proc. ASRU}, 2019, pp. 276--283.

\bibitem{kanda2019guided}
Naoyuki Kanda, Christoph Boeddeker, Jens Heitkaemper, Yusuke Fujita, Shota
  Horiguchi, Kenji Nagamatsu, and Reinhold Haeb-Umbach,
\newblock ``Guided source separation meets a strong {ASR} backend:
  {Hitachi/Paderborn University} joint investigation for dinner party {ASR},''
\newblock in {\em Proc. Interspeech}, 2019, pp. 1248--1252.

\bibitem{medennikov2020stc}
Ivan Medennikov, Maxim Korenevsky, Tatiana Prisyach, Yuri Khokhlov, Mariya
  Korenevskaya, Ivan Sorokin, Tatiana Timofeeva, Anton Mitrofanov, Andrei
  Andrusenko, Ivan Podluzhny, et~al.,
\newblock ``The {STC} system for the {CHiME-6} challenge,''
\newblock in {\em CHiME 2020 Workshop on Speech Processing in Everyday
  Environments}, 2020.

\bibitem{hershey2016deep}
John~R Hershey, Zhuo Chen, Jonathan Le~Roux, and Shinji Watanabe,
\newblock ``Deep clustering: Discriminative embeddings for segmentation and
  separation,''
\newblock in {\em Proc. ICASSP}, 2016, pp. 31--35.

\bibitem{chen2017deep}
Zhuo Chen, Yi~Luo, and Nima Mesgarani,
\newblock ``Deep attractor network for single-microphone speaker separation,''
\newblock in {\em Proc. ICASSP}, 2017, pp. 246--250.

\bibitem{yu2017permutation}
Dong Yu, Morten Kolb{\ae}k, Zheng-Hua Tan, and Jesper Jensen,
\newblock ``Permutation invariant training of deep models for
  speaker-independent multi-talker speech separation,''
\newblock in {\em Proc. ICASSP}. IEEE, 2017, pp. 241--245.

\bibitem{yu2017recognizing}
Dong Yu, Xuankai Chang, and Yanmin Qian,
\newblock ``Recognizing multi-talker speech with permutation invariant
  training,''
\newblock {\em Proc. Interspeech 2017}, pp. 2456--2460, 2017.

\bibitem{seki2018purely}
Hiroshi Seki, Takaaki Hori, Shinji Watanabe, Jonathan Le~Roux, and John~R
  Hershey,
\newblock ``A purely end-to-end system for multi-speaker speech recognition,''
\newblock in {\em Proc. ACL}, 2018, pp. 2620--2630.

\bibitem{chang2019end}
Xuankai Chang, Yanmin Qian, Kai Yu, and Shinji Watanabe,
\newblock ``End-to-end monaural multi-speaker {ASR} system without
  pretraining,''
\newblock in {\em Proc. ICASSP}, 2019, pp. 6256--6260.

\bibitem{chang2019mimo}
Xuankai Chang, Wangyou Zhang, Yanmin Qian, Jonathan~Le Roux, and Shinji
  Watanabe,
\newblock ``{MIMO-SPEECH}: End-to-end multi-channel multi-speaker speech
  recognition,''
\newblock in {\em Proc. ASRU}, 2019, pp. 237--244.

\bibitem{kanda2019acoustic}
Naoyuki Kanda, Yusuke Fujita, Shota Horiguchi, Rintaro Ikeshita, Kenji
  Nagamatsu, and Shinji Watanabe,
\newblock ``Acoustic modeling for distant multi-talker speech recognition with
  single-and multi-channel branches,''
\newblock in {\em Proc. ICASSP}, 2019, pp. 6630--6634.

\bibitem{kanda2019auxiliary}
Naoyuki Kanda, Shota Horiguchi, Ryoichi Takashima, Yusuke Fujita, Kenji
  Nagamatsu, and Shinji Watanabe,
\newblock ``Auxiliary interference speaker loss for target-speaker speech
  recognition,''
\newblock in {\em Proc. Interspeech}, 2019, pp. 236--240.

\bibitem{wang2019speech}
Peidong Wang, Zhuo Chen, Xiong Xiao, Zhong Meng, Takuya Yoshioka, Tianyan Zhou,
  Liang Lu, and Jinyu Li,
\newblock ``Speech separation using speaker inventory,''
\newblock in {\em Proc. ASRU}, 2019, pp. 230--236.

\bibitem{von2019all}
Thilo von Neumann, Keisuke Kinoshita, Marc Delcroix, Shoko Araki, Tomohiro
  Nakatani, and Reinhold Haeb-Umbach,
\newblock ``All-neural online source separation, counting, and diarization for
  meeting analysis,''
\newblock in {\em Proc. ICASSP}, 2019, pp. 91--95.

\bibitem{kinoshita2020tackling}
Keisuke Kinoshita, Marc Delcroix, Shoko Araki, and Tomohiro Nakatani,
\newblock ``Tackling real noisy reverberant meetings with all-neural source
  separation, counting, and diarization system,''
\newblock {\em arXiv preprint arXiv:2003.03987}, 2020.

\bibitem{park2018multimodal}
Tae~Jin Park and Panayiotis Georgiou,
\newblock ``Multimodal speaker segmentation and diarization using lexical and
  acoustic cues via sequence to sequence neural networks,''
\newblock in {\em Proc. Interspeech}, 2018, pp. 1373--1377.

\bibitem{park2020speaker}
Tae~Jin Park, Kyu~J Han, Jing Huang, Xiaodong He, Bowen Zhou, Panayiotis
  Georgiou, and Shrikanth Narayanan,
\newblock ``Speaker diarization with lexical information,''
\newblock in {\em Proc. Interspeech}, 2019, pp. 391--395.

\bibitem{el2019joint}
Laurent El~Shafey, Hagen Soltau, and Izhak Shafran,
\newblock ``Joint speech recognition and speaker diarization via sequence
  transduction,''
\newblock in {\em Proc. Interspeech}, 2019, pp. 396--400.

\bibitem{mao2020speech}
Huanru~Henry Mao, Shuyang Li, Julian McAuley, and Garrison Cottrell,
\newblock ``Speech recognition and multi-speaker diarization of long
  conversations,''
\newblock {\em arXiv preprint arXiv:2005.08072}, 2020.

\bibitem{kanda2019simultaneous}
Naoyuki Kanda, Shota Horiguchi, Yusuke Fujita, Yawen Xue, Kenji Nagamatsu, and
  Shinji Watanabe,
\newblock ``Simultaneous speech recognition and speaker diarization for
  monaural dialogue recordings with target-speaker acoustic models,''
\newblock in {\em Proc. ASRU}, 2019.

\bibitem{kanda2020joint}
Naoyuki Kanda, Yashesh Gaur, Xiaofei Wang, Zhong Meng, Zhuo Chen, Tianyan Zhou,
  and Takuya Yoshioka,
\newblock ``Joint speaker counting, speech recognition, and speaker
  identification for overlapped speech of any number of speakers,''
\newblock {\em arXiv preprint arXiv:2006.10930}, 2020.

\bibitem{chen2020continuous}
Zhuo Chen, Takuya Yoshioka, Liang Lu, Tianyan Zhou, Zhong Meng, Yi~Luo, Jian
  Wu, and Jinyu Li,
\newblock ``Continuous speech separation: dataset and analysis,''
\newblock in {\em Proc. ICASSP}, 2020 (to appear).

\bibitem{variani2014deep}
Ehsan Variani, Xin Lei, Erik McDermott, Ignacio~Lopez Moreno, and Javier
  Gonzalez-Dominguez,
\newblock ``Deep neural networks for small footprint text-dependent speaker
  verification,''
\newblock in {\em Proc. ICASSP}, 2014, pp. 4052--4056.

\bibitem{kanda2020sot}
Naoyuki Kanda, Yashesh Gaur, Xiaofei Wang, Zhong Meng, and Takuya Yoshioka,
\newblock ``Serialized output training for end-to-end overlapped speech
  recognition,''
\newblock {\em arXiv preprint arXiv:2003.12687}, 2020.

\bibitem{park2019auto}
Tae~Jin Park, Kyu~J Han, Manoj Kumar, and Shrikanth Narayanan,
\newblock ``Auto-tuning spectral clustering for speaker diarization using
  normalized maximum eigengap,''
\newblock {\em IEEE Signal Processing Letters}, vol. 27, pp. 381--385, 2019.

\bibitem{von2007tutorial}
Ulrike Von~Luxburg,
\newblock ``A tutorial on spectral clustering,''
\newblock {\em Statistics and computing}, vol. 17, no. 4, pp. 395--416, 2007.

\bibitem{panayotov2015librispeech}
Vassil Panayotov, Guoguo Chen, Daniel Povey, and Sanjeev Khudanpur,
\newblock ``Librispeech: an {ASR} corpus based on public domain audio books,''
\newblock in {\em Proc. ICASSP}, 2015, pp. 5206--5210.

\bibitem{povey2011kaldi}
Daniel Povey, Arnab Ghoshal, Gilles Boulianne, Lukas Burget, Ondrej Glembek,
  Nagendra Goel, Mirko Hannemann, Petr Motlicek, Yanmin Qian, Petr Schwarz,
  et~al.,
\newblock ``The {Kaldi} speech recognition toolkit,''
\newblock in {\em ASRU}, 2011.

\bibitem{ko2017study}
Tom Ko, Vijayaditya Peddinti, Daniel Povey, Michael~L Seltzer, and Sanjeev
  Khudanpur,
\newblock ``A study on data augmentation of reverberant speech for robust
  speech recognition,''
\newblock in {\em Proc. ICASSP}, 2017, pp. 5220--5224.

\bibitem{habets2007generating}
Emanu{\"e}l~AP Habets and Sharon Gannot,
\newblock ``Generating sensor signals in isotropic noise fields,''
\newblock {\em The Journal of the Acoustical Society of America}, vol. 122, no.
  6, pp. 3464--3470, 2007.

\bibitem{nagrani2017voxceleb}
Arsha Nagrani, Joon~Son Chung, and Andrew Zisserman,
\newblock ``Voxceleb: A large-scale speaker identification dataset,''
\newblock in {\em Proc. Interspeech}, 2017, pp. 2616--2620.

\bibitem{chung2018voxceleb2}
Joon~Son Chung, Arsha Nagrani, and Andrew Zisserman,
\newblock ``Voxceleb2: Deep speaker recognition,''
\newblock in {\em Proc. Interspeech}, 2018, pp. 1086--1090.

\bibitem{zhou2019cnn}
Tianyan Zhou, Yong Zhao, Jinyu Li, Yifan Gong, and Jian Wu,
\newblock ``{CNN} with phonetic attention for text-independent speaker
  verification,''
\newblock in {\em Proc. ASRU}, 2019, pp. 718--725.

\bibitem{ba2016layer}
Jimmy~Lei Ba, Jamie~Ryan Kiros, and Geoffrey~E Hinton,
\newblock ``Layer normalization,''
\newblock {\em arXiv preprint arXiv:1607.06450}, 2016.

\bibitem{chorowski2015attention}
Jan~K Chorowski, Dzmitry Bahdanau, Dmitriy Serdyuk, Kyunghyun Cho, and Yoshua
  Bengio,
\newblock ``Attention-based models for speech recognition,''
\newblock in {\em Proc. NIPS}, 2015, pp. 577--585.

\bibitem{kudo2018subword}
Taku Kudo,
\newblock ``Subword regularization: Improving neural network translation models
  with multiple subword candidates,''
\newblock {\em arXiv preprint arXiv:1804.10959}, 2018.

\bibitem{fujita2019end}
Yusuke Fujita, Naoyuki Kanda, Shota Horiguchi, Kenji Nagamatsu, and Shinji
  Watanabe,
\newblock ``End-to-end neural speaker diarization with permutation-free
  objectives,''
\newblock {\em Proc. Interspeech}, pp. 4300--4304, 2019.

\bibitem{fujita2019end2}
Yusuke Fujita, Naoyuki Kanda, Shota Horiguchi, Yawen Xue, Kenji Nagamatsu, and
  Shinji Watanabe,
\newblock ``End-to-end neural speaker diarization with self-attention,''
\newblock in {\em Proc. ASRU}, 2019, pp. 296--303.

\bibitem{medennikov2020target}
Ivan Medennikov, Maxim Korenevsky, Tatiana Prisyach, Yuri Khokhlov, Mariya
  Korenevskaya, Ivan Sorokin, Tatiana Timofeeva, Anton Mitrofanov, Andrei
  Andrusenko, Ivan Podluzhny, et~al.,
\newblock ``Target-speaker voice activity detection: a novel approach for
  multi-speaker diarization in a dinner party scenario,''
\newblock {\em arXiv preprint arXiv:2005.07272}, 2020.

\end{thebibliography}

\end{document}